\documentclass{article}

\usepackage[utf8]{inputenc}
\usepackage[T1]{fontenc}

\PassOptionsToPackage{table, dvipsnames}{xcolor}
\usepackage{xcolor}

\PassOptionsToPackage{colorlinks=true, citecolor=blue, linkcolor=blue, urlcolor=black}{hyperref}

\usepackage{arxiv}
\usepackage[utf8]{inputenc}
\usepackage[T1]{fontenc}
\usepackage{url}
\usepackage{booktabs}
\usepackage{nicefrac}
\usepackage{microtype}
\usepackage{lipsum}
\usepackage{graphicx}
\usepackage{natbib}
\usepackage{doi}
\usepackage{algorithm}
\usepackage{multirow}
\usepackage{multicol}
\usepackage{amsmath,amssymb,amsfonts}
\usepackage{amsthm}
\usepackage{mathrsfs}
\usepackage{float}
\usepackage{algorithmic}
\usepackage{bm}
\usepackage{caption}
\usepackage{hyperref}

\title{Physics-informed conditional diffusion model for generalizable elastic wave-mode separation
}

\author{ \href{https://orcid.org/0000-0001-8868-7967}{\includegraphics[scale=0.06]{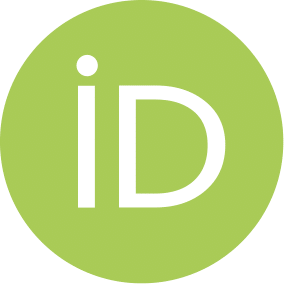}\hspace{1mm}Shijun~Cheng}\\
	Division of Physical Science and Engineering\\
	King Abdullah University of Science and Technology\\
	Thuwal 23955-6900, Saudi Arabia \\
	\texttt{sjcheng.academic@gmail.com} \\
        \And
	\href{https://orcid.org/0000-0002-5950-5201}{\includegraphics[scale=0.06]{orcid.png}\hspace{1mm}Xinru Mu} \\
	Division of Physical Science and Engineering\\
	King Abdullah University of Science and Technology\\
	Thuwal 23955-6900, Saudi Arabia \\
	\texttt{xinru.mu@kaust.edu.sa} \\
        \And
	\href{https://orcid.org/0000-0002-9363-9799}{\includegraphics[scale=0.06]{orcid.png}\hspace{1mm}Tariq~Alkhalifah} \\
	Division of Physical Science and Engineering\\
	King Abdullah University of Science and Technology\\
	Thuwal 23955-6900, Saudi Arabia \\
	\texttt{tariq.alkhalifah@kaust.edu.sa} \\
}

\renewcommand{\shorttitle}{Physics-informed conditional diffusion model for generalizable elastic wave-mode separation}

\hypersetup{
pdftitle={A template for the arxiv style},
pdfsubject={q-bio.NC, q-bio.QM},
pdfauthor={David S.~Hippocampus, Elias D.~Striatum},
pdfkeywords={First keyword, Second keyword, More},
}

\begin{document}
\maketitle

\begin{abstract}
Traditional elastic wavefield separation methods, while accurate, often demand substantial computational resources, especially for large geological models or 3D scenarios. Purely data-driven neural network approaches can be more efficient, but may fail to generalize and maintain physical consistency due to the absence of explicit physical constraints. Here, we propose a physics-informed conditional diffusion model for elastic wavefield separation that seamlessly integrates domain-specific physics equations into both the training and inference stages of the reverse diffusion process. Conditioned on full elastic wavefields and subsurface P- and S-wave velocity profiles, our method directly predicts clean P-wave modes while enforcing Laplacian separation constraints through physics-guided loss and sampling corrections. Numerical experiments on diverse scenarios yield the separation results that closely match conventional numerical solutions but at a reduced cost, confirming the effectiveness and generalizability of our approach.
\end{abstract}

\section{\textbf{Introduction}}
Elastic wavefield separation plays a critical role in seismic imaging, particularly in advanced techniques such as reverse time migration (RTM) \citep{zhang20102d, cheng2014fast}. In elastic media, different wave modes, which primarily include compressional (P) and shear (S) waves, carry distinct subsurface information. Accurate separation of these modes is essential because it not only enhances the resolution and clarity of the final image but also reduces image artifacts and, thus, improves the reliability of subsequent interpretation processes. 

Over the years, several classical methods have been developed for elastic wavefield separation. One widely used approach is the Helmholtz decomposition, which leverages the divergence and curl operators to isolate the P- and S-wave components \citep{aki2002quantitative, yan2009elastic}. In addition, techniques based on the wavenumber domain, such as Fourier-domain filtering, exploit the distinct spectral characteristics of different wave modes to achieve separation \citep{dellinger1990wave, zhang20102d, yan2011improving, cheng2016simulating, zhu2017elastic, yang2019elastic, mu2024attenuation}. Although these approaches have proven effective in many settings, they often require high computational cost, especially for anisotropic media and, also, can be sensitive to noise. 

In recent years, data-driven techniques have emerged as promising alternatives for elastic wavefield separation. Neural network (NN)–based methods have shown the potential to learn complex relationships from data and can significantly improve separation performance while reducing computational overhead. For example, \cite{wang2019ps} utilized convolutional neural networks (CNNs) to learn optimal filters for P/S decomposition in isotropic elastic wavefields, effectively extracting P- and S-wave modes. Similarly, \cite{kaur2021fast} introduced a fast generative adversarial network-based algorithm to achieve elastic wave-mode separation, leveraging adversarial training to enhance separation performance. \cite{wei2021deep} presented an efficient training dataset construction strategy based on kinematic and dynamic feature analysis and, by integrating it with a deep convolutional NN (CNN), achieved automatic P- and S-wave separation for multicomponent vertical seismic profiling (VSP) data. \cite{huang2023p} also applied CNN to separate P/S waves from multicomponent seismic land data, addressing challenges posed by near-surface complexities. \cite{meng2025up} framed the upgoing/downgoing separation and P/S-wave decomposition of distributed acoustic sensing-VSP data as a multi-task supervised learning problem, generated extensive labels from expanded elastic models, and jointly trained a multi-task CNN with shared representations. Despite these advances, a common limitation of these approaches is that they are almost entirely data-driven. They rely on learning from examples without explicitly incorporating physical principles into the models. This lack of physical guidance can lead to generalization issues. As a result, models trained on a specific dataset may not perform well when confronted with data from different geological settings. 

To address the generalization gap of purely data-driven schemes, \cite{mu2025separationpinn} recently introduced SeparationPINN, a self-supervised P/S-wave separation framework that embeds the elastic wave separation equations \citep{zhu2017elastic} directly into the loss function of a physics-informed neural network. By predicting the P- and S-wave mode with two separate fully connected networks and enforcing both PDE residuals and data-consistency boundary constraints, SeparationPINN delivers mesh-free separation that remains accurate in both homogeneous and heterogeneous models. However, this accuracy comes at a steep price: the network must be retrained from scratch for each time snapshot. While one can exploit transfer learning by initializing a new snapshot’s PINN with weights from a previous snapshot to speed convergence, the need to perform a complete PINN training for every time step still requires a high computational cost for large-scale seismic applications. 

To overcome this bottleneck, we propose a physics-informed conditional diffusion model (PICDM) for elastic wavefield separation that learns once, applies everywhere. The beauty of our framework lies in its integration of physical principles directly into both the training and inference processes. During training, we incorporate the elastic wavefield separation equations to formulate a physics-informed loss, which constrains the diffusion model to learn representations that correspond to known physical behaviors. At inference time, this same physics loss guides the sampling process, ensuring that the separated wavefields satisfy the underlying physical criteria. We validate the efficacy and robust generalization of our approach through extensive tests on multiple velocity models, demonstrating superior separation performance even in complex geological settings.

\section{\textbf{Method}}
In this section, we first presents the governing first‑order velocity–stress formulation of elastic wave propagation and the classical decoupling approach for extracting P‑ and S‑wave modes. Building on this theoretical foundation, we then formulate elastic wave mode separation as a conditional generative task and detail our conditional diffusion model (CDM). Furthermore, we show how we embed physics‑based constraints not only in the training objective but also at each inference step to ensure that the separated wavefields remain physically plausible. Finally, a detail description of our network architecture follows.

\subsection{Theoretical basis of elastic wave mode separation}\label{subsec:decoupling}
In a 2D isotropic elastic medium, wave propagation can be described by the first-order velocity-stress elastic wave equations, given by \citep{virieux1984sh, virieux1986p}:
\begin{equation}\label{eq1}
    \begin{aligned}
    \rho\,\partial_t V_x &= \partial_x \sigma_{xx} + \partial_z \sigma_{xz} + f_x, \\
    \rho\,\partial_t V_z &= \partial_x \sigma_{xz} + \partial_z \sigma_{zz} + f_z, \\
    \partial_t \sigma_{xx} &= (\lambda+2\mu)\,\partial_x V_x + \lambda\,\partial_z V_z, \\
    \partial_t \sigma_{zz} &= \lambda\,\partial_x V_x + (\lambda+2\mu)\,\partial_z V_z, \\
    \partial_t \sigma_{xz} &= \mu\,(\partial_z V_x + \partial_x V_z),
   \end{aligned}
\end{equation}
where $V_x$ and $V_z$ are the horizontal and vertical particle velocities, respectively, $\sigma_{xx}$, $\sigma_{zz}$, and $\sigma_{xz}$ are stress tensors, 
$\rho$ denotes the density, $\lambda$ and $\mu$ are Lamé’s elastic parameters, $f_x$ and $f_z$ represent the external source terms, and $\partial_{i}$ are the partial derivatives with respect to the variables ${i}=x$, ${i}=z$, and $i=t$, respectively. Since the particle velocities and stresses hold the coupled nature, the solutions of these equations inherently include both P- and S-wave modes. 

To effectively decouple seismic wavefields, a popular approach was proposed by \cite{zhu2017elastic}, which explicitly separates elastic wavefields into their P- and S-wave modes by applying spatial derivatives and the Laplacian operator. Specifically, the elastic velocity wavefield $\mathbf{V}=(V_x,V_z)$ can be decomposed into P- $(V_x^p,V_z^p)$ and S-wave $(V_x^s,V_z^s)$ modes via:
\begin{equation}\label{eq2}
  \begin{aligned}
    \Delta{V_{x}^{p}} &= \bigl(\partial^{2}_x V_x + \partial_x\partial_z V_z\bigr), 
    &\Delta{V_{z}^{p}} &= \bigl(\partial_x\partial_z V_x + \partial^{2}_z V_z\bigr),\\
    \Delta{V_{x}^{s}} &= \bigl(\partial^{2}_z V_x - \partial_x\partial_z V_z\bigr), 
    &\Delta{V_{z}^{s}} &= \bigl(\partial^{2}_x V_z - \partial_x\partial_z V_x\bigr),
  \end{aligned}
\end{equation}
where \(\Delta=\partial_x^2+\partial_z^2\), and $\partial_{i}^m$ are the $m$-order spatial partial derivatives with respect to the variables ${i}=x$ or ${i}=z$. 

Although Equation \ref{eq2} provides a straightforward way to separate elastic wavefields, it involves high-order mixed derivatives that are typically evaluated by pseudo-spectral methods, leading to substantial cost. The situation worsens in anisotropic media, where the separation operator becomes even more complicated.
These complexities motivate a learned alternative
that can reduce the computational burden while retaining physical fidelity. 

\subsection{Elastic wavefield separation via conditional diffusion models}
To address the limitations of conventional ML methods, we frame the elastic wavefield separation problem as a generative modeling task and propose to employ a CDM for P- and S-wave mode separation. Given that the elastic wavefield is inherently a combination of the P- and S-wave modes, we primarily focus on directly separating the P-wave mode $(V_x^{p},V_z^{p})$. Subsequently, the S-wave mode can be straightforwardly obtained by subtracting the predicted P-wave mode from the original elastic wavefield. 

\begin{figure}[!htb]
\centering
\includegraphics[width=0.95\textwidth]{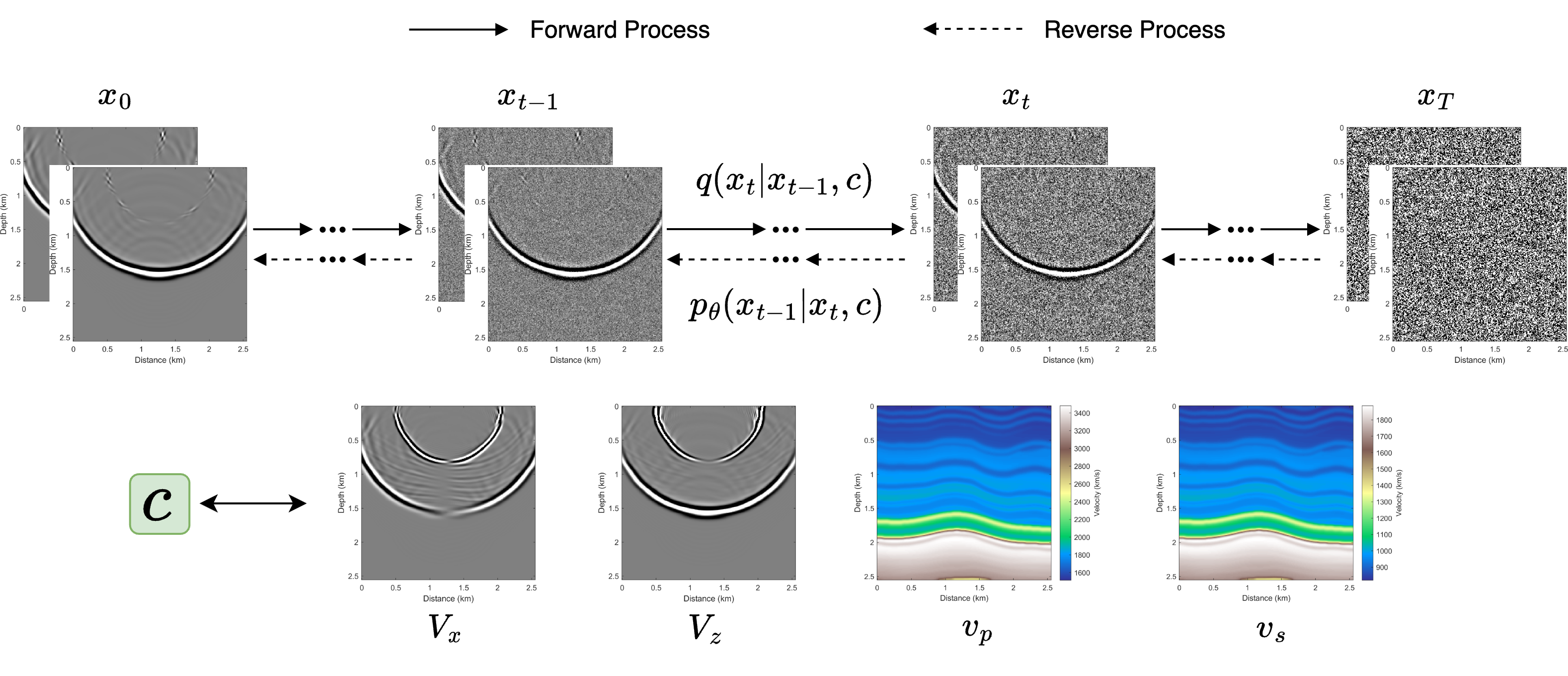}
\caption{Schematic of our conditional diffusion model for P-wave mode separation. Forward diffusion (solid arrow) progressively corrupts the clean P-wave mode \(x_0=(V_x^p,V_z^p)\) with Gaussian noise over \(T\) steps, yielding \(x_0, \dots, x_T\). Reverse diffusion uses a neural network \(f_\theta\) conditioned on \(\mathbf{c}=(V_x,V_z,v_p,v_s)\) (dashed arrow) to denoise from \(x_T\) back to \(x_0\).}
\label{fig1}
\end{figure}

To achieve this, we adopt a CDM framework consisting of two key processes: the forward diffusion and the reverse diffusion processes, which are illustrated in Figure \ref{fig1}. The forward diffusion process progressively corrupts the clean P-wave mode $x_0=(V_x^{p},V_z^{p})$ with Gaussian noise across $T$ time steps. By the final step, the signal becomes almost pure Gaussian noise \citep{ho2020denoising}. Formally,
\begin{equation}\label{eq3}
    q(x_t \mid x_{t-1}) 
    = \mathcal{N}\!\bigl(x_t;\, \sqrt{\alpha_t}\,x_{t-1},\, (1-\alpha_t)\mathbf{I}\bigr),
\end{equation}
where $x_t$ denotes the noisy P-wave mode at step $t$, and $\alpha_t$ controls the noise schedule. Equivalently, one can sample \(x_t\) in closed form via the reparameterization trick:
\begin{equation}\label{eq4}
  x_t
  = \sqrt{\bar\alpha_t}\,x_0
  \;+\;
  \sqrt{1-\bar\alpha_t}\,\epsilon,
  \quad
  \epsilon\sim\mathcal{N}(0,\mathbf{I}),
  \quad
  \bar\alpha_t=\prod_{s=1}^t\alpha_s.
\end{equation}
This formula allows us to directly generate any noisy intermediate \(x_t\) (for arbitrary \(t\)) given the clean wavefield \(x_0\) and the sampled Gaussian noise \(\epsilon\), which is crucial for constructing our training pairs \(\{x_t,\mathbf{c},t;x_0\}\), as we will introduce in the following subsection.

Conversely, the reverse diffusion process seeks to recover the clean P-wave mode $x_0$ from pure noise. Starting from $x_T \sim \mathcal{N}(0,\,\mathbf{I})$, we iteratively apply
\begin{equation}\label{eq5}
  p_\theta(x_{t-1}\!\mid\!x_t,\mathbf{c})
  = \mathcal{N}\!\Bigl(x_{t-1};\,
    \boldsymbol\mu_\theta(x_t,t,\mathbf{c}),\,
    \Sigma_t\Bigr),
\end{equation}
where
\begin{equation}\label{eq6}
\mathbf{c} = \bigl(V_x,V_z,v_p,v_s\bigr)
\end{equation}
denotes the conditioning inputs: the original elastic wavefield components $(V_x,V_z)$ and the subsurface P- and S-wave velocity models $(v_p,v_s)$. We will justify, in the discussion section, why having the velocities as additional conditions is preferred, though they do not appear in the separation Equation \ref{eq2}.  
The mean $\boldsymbol\mu_\theta$ is predicted by a trained neural network $f_\theta$, typically a U-Net augmented with timestep embeddings and several layers to incorporate $\mathbf{c}$. 

Therefore, the forward process (Equation~\ref{eq3}) corrupts the clean P-wave mode into near Gaussian noise over $T$ steps, while the reverse process (Equation~\ref{eq5}) learns to invert this corruption, producing a high-fidelity estimate of $(V_x^p,V_z^p)$ from pure Gaussian noise under the guidance of the original elastic wavefield and velocity models.  

\subsection{Physics-informed training of the reverse denoiser}\label{subsec:training}

Building on the reverse diffusion formulation in Equation ~\ref{eq5}, we should train a denoising network \(f_\theta\) to estimate the mean $\boldsymbol\mu_\theta$ from a noisy input \(x_t\). Here, we adopt an $x_0$-prediction parameterization to better incorporate the physics-based constraints. That is, the network directly predicts \(\hat{x}_{0,\theta}\), i.e., the clean P-wave mode \((V_x^p,V_z^p)\), rather than the noise added at diffusion step $t$. We then reconstruct the reverse mean via:
\begin{equation}\label{eq7}
  \boldsymbol\mu_\theta(x_t,t,\mathbf{c}) =
  \frac{\sqrt{\bar\alpha_{t-1}}\,\beta_t}{1-\bar\alpha_t}\,x_t +
  \frac{\sqrt{\alpha_t}(1-\bar\alpha_{t-1})}{1-\bar\alpha_t}\,\hat{x}_{0,\theta},
  \quad \beta_t=1-\alpha_t.
\end{equation}

To prepare training examples, we use the reparameterization diffusion equation~\ref{eq4} to corrupt each ground-truth P-wave mode \((V_x^p,V_z^p)\) into noisy samples \(x_t\) of it, with noise depending on the time step $t$. Each noisy sample, along with conditional
inputs \(\mathbf{c}=\bigl(V_x, V_z, v_p, v_s\bigr)\), form the network’s input, while the clean P-wave mode
\(\bigl(V_x^p, V_z^p\bigr)\) serves as the target output. 

Once we have constructed these input-target pairs
$\{\textbf{input:}~x_t, \mathbf{c}, t; \textbf{target:}~V_x^p, V_z^p\}$, our
goal is to train the network to map the noisy P-wave mode
\(x_t\) (conditioned on \(\mathbf{c}\)) to the clean P-wave mode \(x_0=\bigl(V_x^p, V_z^p\bigr)\). Specifically, we adopt two loss terms: 
\begin{enumerate}
    \item \textbf{Data loss.} We first measure how closely the network output \(\bigl(\hat{V}_x^p, \hat{V}_z^p\bigr)\) matches the ground-truth
    \(\bigl(V_x^p, V_z^p\bigr)\) in an \(\ell_2\)-sense:
    \begin{equation}\label{eq8}
      \mathcal{L}_{\mathrm{data}} 
      \;=\;
      \bigl\lVert \hat{V}_x^p - V_x^p \bigr\rVert_2^2
      \;+\;
      \bigl\lVert \hat{V}_z^p - V_z^p \bigr\rVert_2^2.
    \end{equation}
    This loss term ensures that the predicted P-wave mode remains close to the reference data provided by the numerical solver (e.g., Equation~\ref{eq2}).

    \item \textbf{Physics-informed loss.} In addition to matching the ground truth, we incorporate the
    physical constraints derived from Zhu’s separation equations (Equation~\ref{eq2}). Concretely, we require the network’s output
    \(\bigl(\hat{V}_x^p, \hat{V}_z^p\bigr)\) to satisfy the Laplacian
    relationships for P-wave separation. Formally,
    \begin{equation}\label{eq9}
      \begin{aligned}
      \mathcal{L}_{\mathrm{phys}}
      \;=\;
      \bigl\lVert \Delta\hat{V}_x^p - (\partial^{2}_x{V_x}+\partial_x\partial_z{V_z}) \bigr\rVert_2^2+
      \bigl\lVert \Delta\hat{V}_z^p - (\partial_x\partial_z{V_x}+\partial^{2}_z{V_z})\bigr\rVert_2^2.
      \end{aligned}
    \end{equation}
    By explicitly encouraging the predicted P-wave mode to align with these decoupled operators, we embed domain knowledge about wave propagation into the training process.
\end{enumerate}

Finally, we combine these two losses into a single training objective:
\begin{equation}\label{eq10}
  \mathcal{L}_{\mathrm{train}}
  \;=\;
  \mathcal{L}_{\mathrm{data}}
  \;+\;
  \lambda\,\mathcal{L}_{\mathrm{phys}},
\end{equation}
where \(\lambda\) is a hyperparameter that balances data fidelity against physical consistency. This approach allows us to directly predict \(\bigl(\hat{V}_x^p, \hat{V}_z^p\bigr)\) while simultaneously
incorporating both data-driven and physics-based
constraints, ultimately improving the robustness and accuracy of our P-wave mode separation.

\subsection{Physics-guided P-wave mode generation}
\label{sec:pg_sampling}
After training, we utilize the learned diffusion model to reconstruct the clean P-wave mode from Gaussian noise in a step-by-step reverse diffusion manner. Let \(x_T\) be a sample drawn from a standard Gaussian distribution \(\mathcal{N}(0,\mathbf{I})\).
We then iteratively refine \(x_T \rightarrow x_{T-1} \rightarrow \dots \rightarrow x_{0}\) via the model’s reverse process, conditioned on \(\mathbf{c}\). At each
reverse step \(t\), we perform the following two steps in sequence:
\begin{enumerate}
    \item \textbf{Denoising update:} 
    The network predicts an estimate \(\hat{x}_{t-1}\) given \(x_t\) and the conditional inputs $\mathbf{c}$. This step approximates
    the reverse transition
    \(p_\theta(x_{t-1}\mid x_t, \mathbf{c})\),
    thereby removing a portion of the noise and guiding the sample toward the underlying clean P-wave fields.

    \item \textbf{Physics correction:} 
    The intermediate wavefield may still violate Equation~\eqref{eq2}. Therefore, we further apply a physics-guided correction to the intermediate prediction
    \(\hat{x}_{t-1} = \bigl(\hat{V}_x^p, \hat{V}_z^p\bigr)\)
    by evaluating the physics-informed loss
    \(\mathcal{L}_{\mathrm{phys}}\) (e.g., Equation~\ref{eq9}) and backpropagating its gradient:
    \begin{equation}\label{eq11}
      \bigl(\hat{V}_x^p, \hat{V}_z^p\bigr)
      \;\leftarrow\;
      \bigl(\hat{V}_x^p, \hat{V}_z^p\bigr)
      \;-\;
      \eta \,\nabla \mathcal{L}_{\mathrm{phys}},
    \end{equation}
    where \(\eta\) is a step size that controls the magnitude of the correction. This guides the sampled wave modes to be consistent with the decoupled equations (Equation~\ref{eq2}), thereby maintaining physical plausibility throughout the reverse diffusion.
\end{enumerate}

Repeating these two steps from \(t = T\) down to \(t = 0\) yields a final estimate of the P-wave mode,
\(\bigl(\hat{V}_x^p, \hat{V}_z^p\bigr)\). Once the P-wave mode is recovered, the S-wave mode follows directly by subtraction:
\begin{equation}\label{eq12}
  \bigl(\hat{V}_x^s, \hat{V}_z^s\bigr)
  \;=\;
  \bigl(V_x, V_z\bigr) 
  \;-\;
  \bigl(\hat{V}_x^p, \hat{V}_z^p\bigr).
\end{equation}
Hence, through our PICDM framework, guided by both data-driven and physics-based constraints, we obtain robust separation of elastic wavefields into their P- and S-wave modes.

\subsection{Network architecture}\label{sec:network}
Our denoising network \(f_{\theta}\) is built upon a five-scale U-Net backbone from the publicly available improved denoising diffusion probabilistic models implementation \citep{nichol2021improved}. The network accepts a six-channel input, comprising the noisy P-wave mode \(x_t\), the two elastic velocity components \((V_x, V_z)\), and the two subsurface velocity models \((v_p, v_s)\), and produces a two-channel output corresponding to the clean P-wave mode \((\hat V_x^p, \hat V_z^p)\). The U-Net encoder uses a base channel width of 64 and successively doubles the number of feature maps at each downsampling stage, resulting in channel widths of 64, 128, 256, 512, and 1024. Each scale contains two residual blocks for local feature extraction. To capture long-range dependencies, multi-head self-attention (4 heads) is applied at intermediate feature map resolutions. Scale-shift normalization is used throughout, and stochastic dropout is disabled to maintain deterministic reconstruction. The decoder mirrors the encoder in reverse: at each of the four upsampling stages, feature maps are upscaled by a factor of 2 (via learned transposed convolutions), concatenated with the corresponding encoder features via skip connections, and processed by two residual blocks. Self-attention blocks are again inserted at the same intermediate resolutions to refine nonlocal information during upsampling.  Finally, a \(3\times3\) convolution projects the full-resolution feature map to the two-channel P-wave output \((\hat V_x^p, \hat V_z^p)\).

\section{\textbf{Numerical examples}}
In the following, we conduct a series of numerical experiments to comprehensively assess the effectiveness and generalization capacity of our proposed PIGDM framework. We first detail the construction of the training dataset, as well as all training and implementation settings. Then, we evaluate our method beginning with a simple homogeneous medium. Here, we specifically investigate the impact of physics-informed constraints on wave mode separation fidelity and explore various reverse diffusion step counts to balance computational cost against separation quality. Next, we extend our evaluations to more complex in-distribution scenarios using the SEAM Arid and Overthrust velocity models. Furthermore, we demonstrate the robustness of our approach by applying it to velocity configurations beyond the training distribution. Finally, we  assess scalability on a larger model and, also, investigate temporal extrapolation by applying our model to snapshots beyond the training time window.

\subsection{Training dataset and configuration}
In order to verify the effectiveness of our proposed method, we first build a dataset of elastic models with various levels of complexity using industrial 2D velocity models \citep{billette20052004} and 3D velocity volumes \citep{aminzadeh19963, naranjo2011survey}. We extract 6000 P-wave velocity models, each with a size of \(256 \times 256\) grid points with a grid spacing of 10~m (both $x$ and $z$ directions). For each P-wave velocity model, we obtain the corresponding S-wave velocity model by multiplying each point by a random scaling factor between 0.52 and 0.7. We set the density to a constant value of \(1000 \,\text{kg/m}^3\). 

Next, we use a 10th-order staggered-grid finite-difference scheme to solve the first-order velocity-stress equations and simulate the horizontal ($x$) and vertical ($z$) components of the elastic velocity wavefield \citep{virieux1984sh, virieux1986p}. The source term is loaded on the $z$-direction particle velocity with a Ricker wavelet of peak frequency of 12~Hz. The time step is 1~ms, and the total simulation time is 1.5~s. During the simulation, convolutional perfectly matched layer boundary conditions \citep{collino2001application} surround the computational domain to suppress boundary reflections. To enrich the variety of source locations, we place 5 shots in each model: one shot at a random grid point on the top surface and four shots randomly inside the model. For each shot, we randomly select 5 different time snapshots of the velocity vector wavefield, leading to \(6000 \times 5 \times 5 = 150000\) data examples in our training set. The separated P wave components used for training are obtained using the method developed by \cite{zhu2017elastic}. 

During training, the model learns over 1000 diffusion timesteps with a cosine noise schedule. We train our diffusion model for 100000 iterations using the AdamW optimizer \citep{loshchilov2017decoupled} with a fixed learning rate of \(1 \times 10^{-4}\), a batch size of 16. We also maintain an exponential moving average of the model parameters with a decay rate of 0.999 to stabilize training. The training is conducted on an NVIDIA A100 GPU, requiring approximately 15 hours. 

We emphasis that, unless otherwise specified, all test wavefields presented in the subsequent experiments share the same simulation parameters, i.e., a 12 Hz Ricker wavelet source and 10 m grid spacing. Meanwhile, all test models share the same size of \(256 \times 256\) grid points, matching the training dataset dimensions. More importantly, after training, our model is applied directly to all subsequent tests without any further retraining or fine-tuning.

\begin{figure*}[htbp]
\centering
\includegraphics[width=1\textwidth]{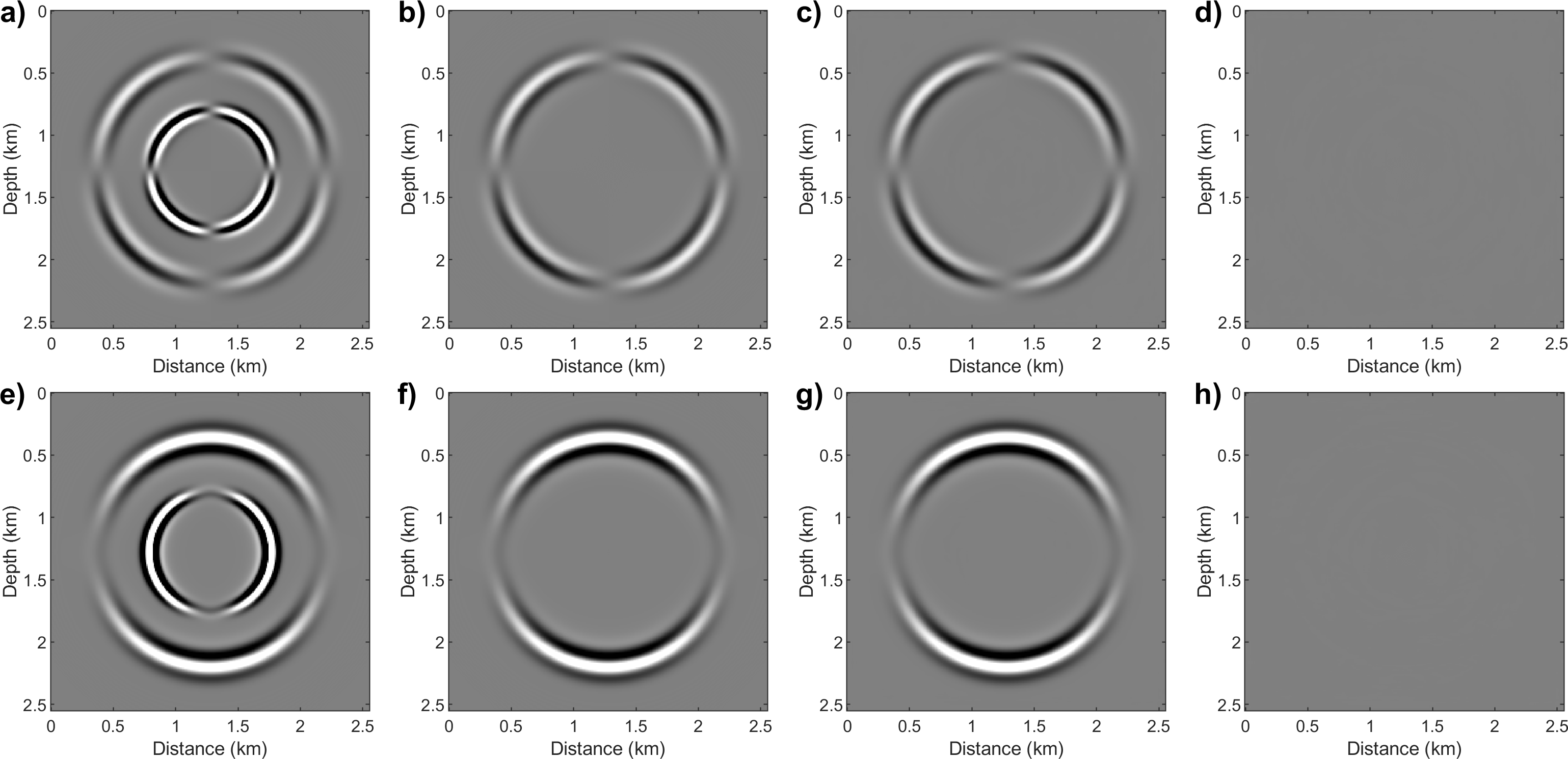}
\caption{Comparison of snapshots at $t=0.4\,\mathrm{s}$ of the P-wave mode separation results for a source at the center of the homogeneous model. The original (a) horizontal and (e) vertical elastic wavefields. The separated P-wave mode in the horizontal (b) and vertical (f) directions obtained using the traditional numerical method. The separated P-wave mode in the horizontal (c) and vertical (g) directions obtained using
the our method. (d) and (h) The differences between the separated P-wave modes shown in second and third column. All wavefields and the differences are plotted at the same scale.}
\label{fig2}
\end{figure*}

\subsection{Homogeneous media}
We begin by evaluating our proposed PICDM method on a simple homogeneous medium with a constant P-wave velocity of \(3 \,\text{km/s}\) and an S-wave velocity of \(1.6 \,\text{km/s}\). The seismic source is positioned at the center of the model, and we select the elastic velocity wavefield components $V_x$ and $V_z$ at the time snapshot of 0.4 s for evaluation. Figure \ref{fig2} displays the separation results obtained by our method. The first and second rows represent the $x$ and $z$ components of the velocity wavefield, respectively. Each row from left to right depicts the original elastic wavefield, the reference P-wave mode computed numerically, the P-wave mode separated by our method, and the residual between the numerical and predicted P-wave modes. Here, we use the denoising diffusion implicit model (DDIM) \citep{song2020denoising} to reduce the sampling from 1000 steps to 50 steps to accelerate the inference process. We can see that our method accurately matches the numerical solution, effectively extracting the P-wave modes for both velocity components. 

\begin{figure*}[htbp]
\centering
\includegraphics[width=0.4\textwidth]{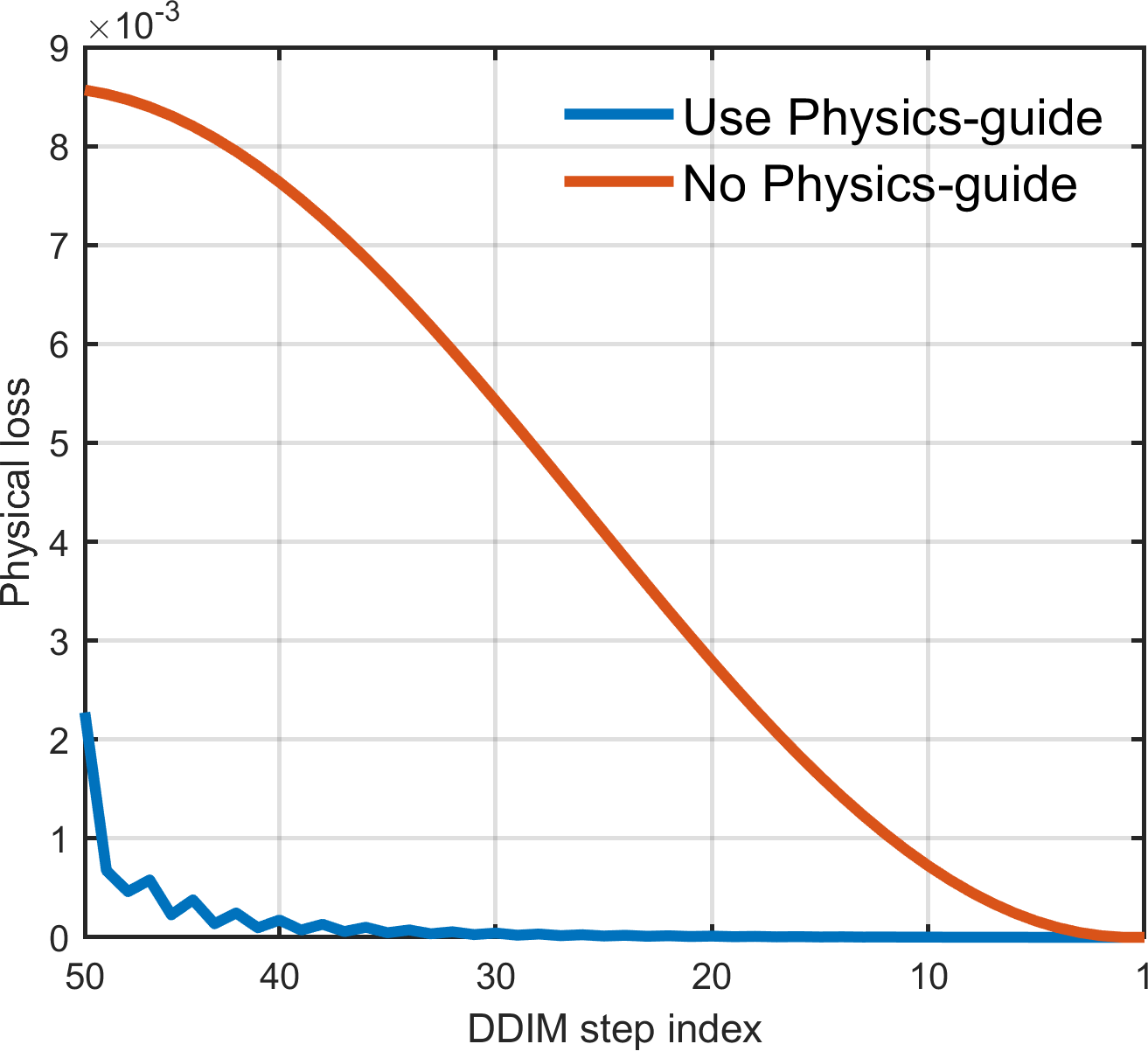}
\caption{Comparison of physics loss $\mathcal{L}_{\mathrm{phys}}$ during the DDIM sampling process for cases without (orange) and with (blue) physics guidance.}
\label{fig3}
\end{figure*}

To further investigate how physics-guided sampling affects the inference process, Figure \ref{fig3} compares the physical loss over reverse sampling steps, with and without physics guidance. We can find that the diffusion model with physics guidance converges to a very low physical residual in far fewer steps. By contrast, the model without physics guidance needs many more steps to reach a comparable accuracy. This suggests that we can reduce the sampling time even more through physics-guided constraints. 

\begin{figure*}[htbp]
\centering
\includegraphics[width=1\textwidth]{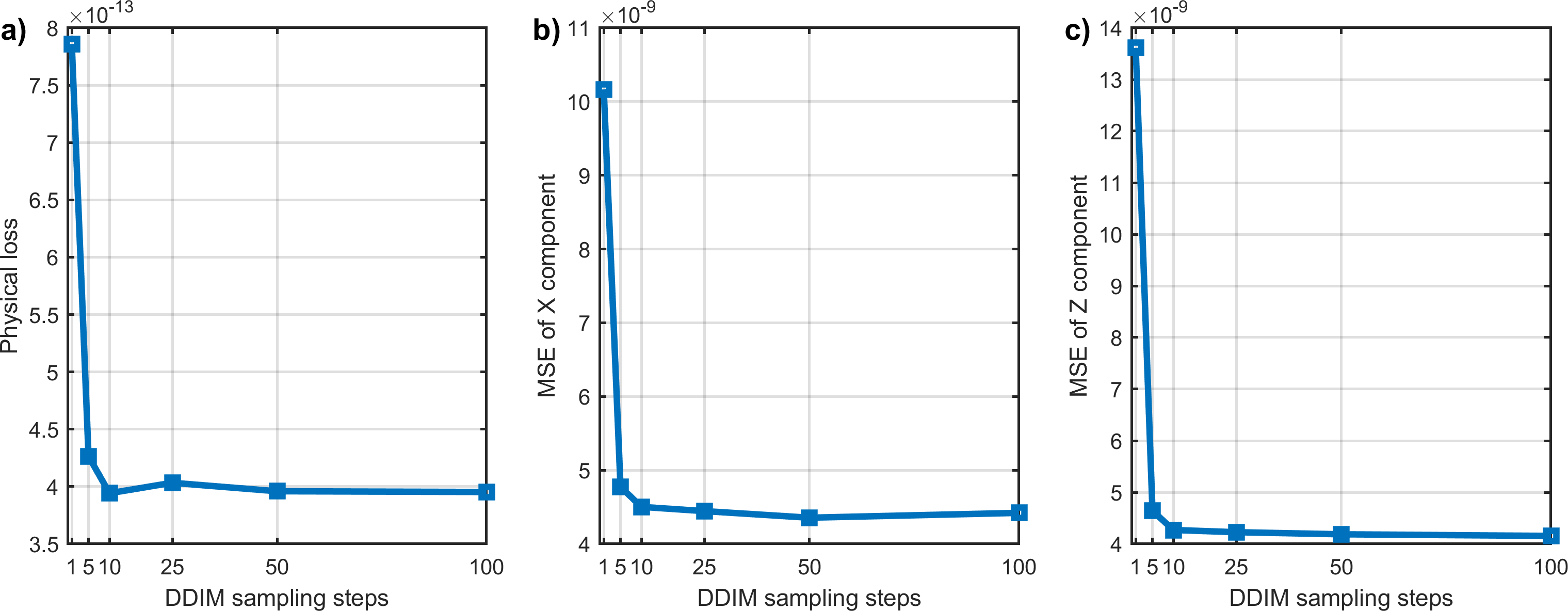}
\caption{The impact of DDIM sampling step count (1, 5, 10, 25, 50, and 100 steps) on wavefield separation quality. (a) Physical loss. The accuracy of the separated P-wave mode for the (b) horizontal and (c) vertical components.}
\label{fig4}
\end{figure*}

Therefore, we conduct an additional analysis to determine the influence of varying sampling step counts on the final separation quality. Specifically, we apply the DDIM sampling approach with step counts of 1, 5, 10, 25, 50, and 100. Figure \ref{fig4} summarizes the corresponding physical losses and separation accuracies, where the accuracy metric is obtained by calculating the MSE between our prediction results and the numerical reference solutions. Panel (a) corresponds to the physical losses, while panels (b) and (c) show the accuracies of the separated P-wave modes for the $x$ and $z$ components, respectively. We can see that, the scenario with only one sampling step exhibits the highest physical loss and errors. Both metrics reduce significantly as the sampling steps increase, with the results stabilizing from 10 steps onward. Based on these observations, we adopt 10 sampling steps combined with physics guidance for subsequent tests to achieve optimal balance between efficiency and accuracy. 

\subsection{SEAM Arid model}
\begin{figure*}[htbp]
\centering
\includegraphics[width=1\textwidth]{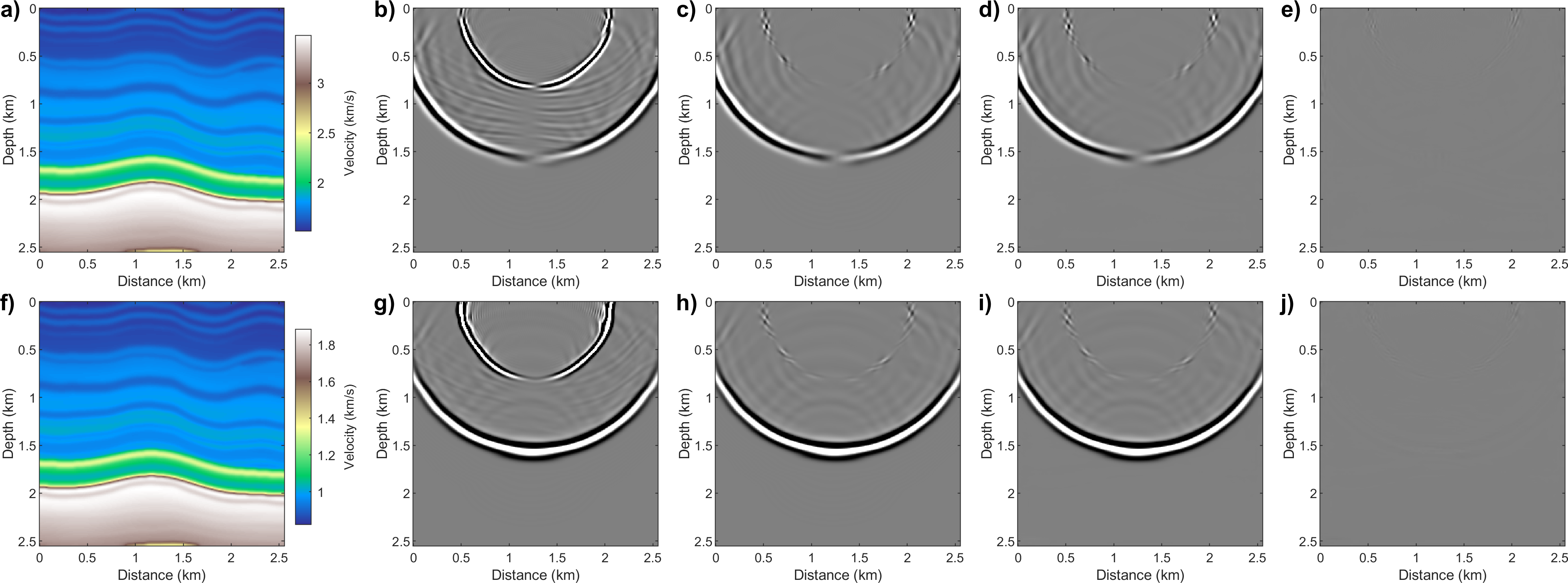}
\caption{Comparison of snapshots at $t=1\,\mathrm{s}$ of the P-wave separation results for a source in the middle of the top surface in the SEAM Arid model. (a)~P- and (f)~S-wave velocities of the layered profiles extracted from SEAM Arid model. The original (b) horizontal and (g) vertical elastic wavefields. The separated P-wave mode in the horizontal (c) and vertical (h) directions obtained using
the traditional numerical method. The separated P-wave mode in the horizontal (d) and vertical (i) directions obtained using the our method. (e) and (j) The differences between the separated P-wave modes shown in third and fourth column. All wavefields and the differences are plotted at the same scale.}
\label{fig5}
\end{figure*}

Next, we test our PICDM method on a more complex in-distribution scenario using the SEAM Arid velocity model \citep{oristaglio2015seam}. We select a layered region from the SEAM Arid model that is excluded from the training dataset, obtaining a distinct P-wave velocity profile. The corresponding S-wave velocity model is computed by multiplying the P-wave velocities by a constant factor of 0.54. Figures \ref{fig5}a and \ref{fig5}f present the extracted layered profiles for P-wave and S-wave velocities, respectively. We place the source at the center of the top surface, and the elastic wavefield at 1.0 s is used for subsequent P-wave mode separation testing. 

Figure~\ref{fig5}, from columns 2 to 5, compares the P-wave modes separated by a traditional numerical method and our method. The first row corresponds to the $x$-component of the velocity wavefield, and the second row corresponds to the $z$-component. From left to right in each row, we show the original elastic wavefield, the solution from the traditional numerical approach, the P-wave field predicted by our model, and the residual between the traditional result and our prediction. We can see that our model’s predictions almost perfectly match the traditional solution, and the residual plots show no noticeable discrepancy. This happened at a much lower cost of the inference compared to the numerical solution. 

\begin{figure*}[htbp]
\centering
\includegraphics[width=1\textwidth]{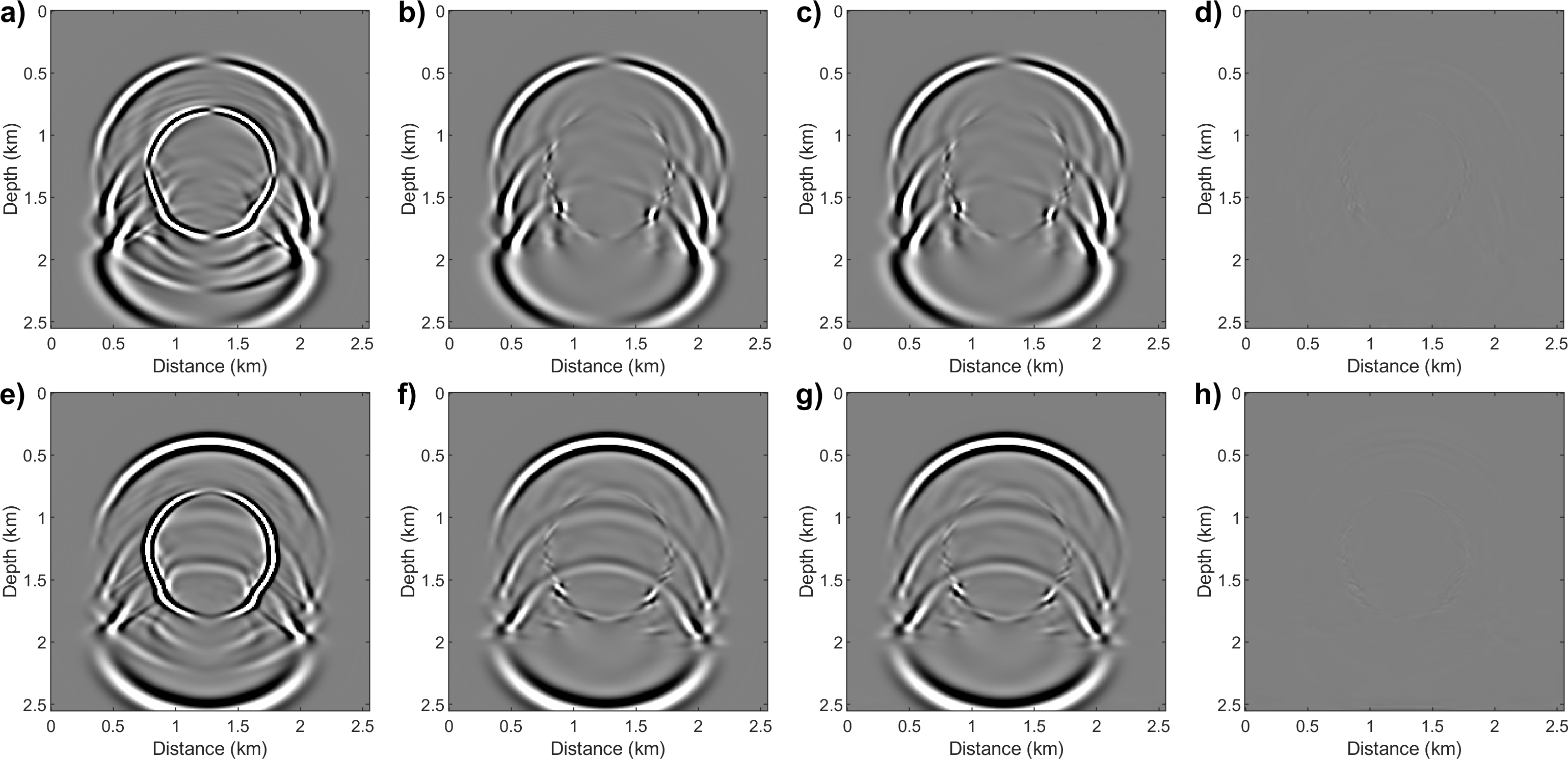}
\caption{Separation performance for SEAM Arid model with the source placed at the center of the model at $t=0.6\,\mathrm{s}$. Except for velocity models, the arrangement of other panels is the same as in Figure~\ref{fig5}.}
\label{fig6}
\end{figure*}

We also evaluate our method's robustness to source position. We still use the same layered model but move the source to the center of the medium. We then record the wavefield snapshot at 0.6~s and separate the P-wave mode. Figure~\ref{fig6} shows the results, following the same panel layout as in Figure~\ref{fig6}, except for velocity model panels. We again observe that our separated P-wave field matches the traditional result closely, and the residual plot shows minimal leakage, indicating strong robustness even for different source locations.

\subsection{Overthrust model}
\begin{figure*}[htbp]
\centering
\includegraphics[width=1\textwidth]{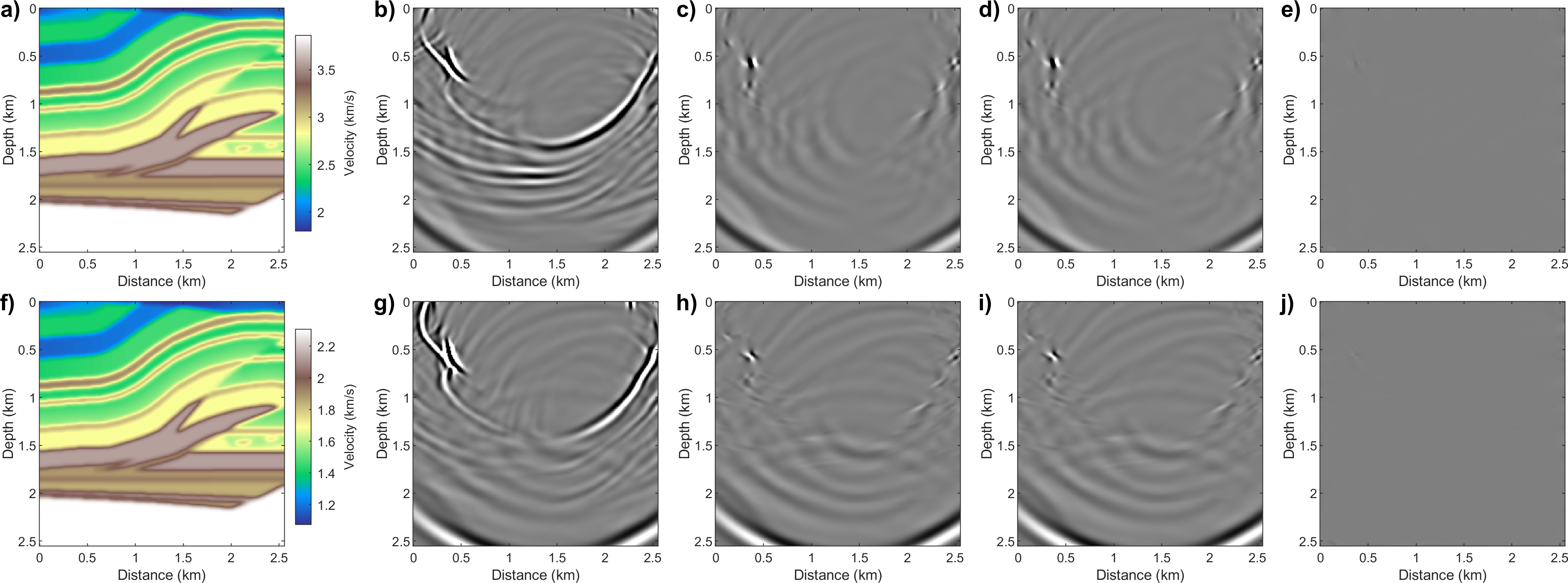}
\caption{Similar with Figure~\ref{fig5}, but for Overthrust model. }
\label{fig7}
\end{figure*}

We further assess our PICDM framework using the Overthrust velocity model \citep{aminzadeh19963}, which includes a more complex geological structures with fault systems. Similar to the previous experiment, we extract a 2D P-wave velocity profile (see Figure~\ref{fig7}a) from part of a 3D overthrust model, which is excluded from the training dataset. The S-wave velocity (see Figure~\ref{fig7}f) is obtained by multiplying the P-wave velocity by~0.6. In this test, we also investigate two scenarios regarding the source locations to validate our method’s adaptability. 

On the one hand, we position the source at the center of the top surface of the Overthrust model and show the elastic wavefield at a time snapshot of 1.0 s. Figure~\ref{fig7} illustrates our separation results. Except for velocity models, panels from left to right present the original elastic wavefields, numerically computed P-wave modes, our PICDM-separated P-wave modes, and residuals between numerical and predicted results, respectively. Even in this more challenging scenario, our method still achieves good agreement with the numerical method, clearly capturing P-wave patterns from complex elastic wavefield with minimal residual discrepancies. 

\begin{figure*}[htbp]
\centering
\includegraphics[width=1\textwidth]{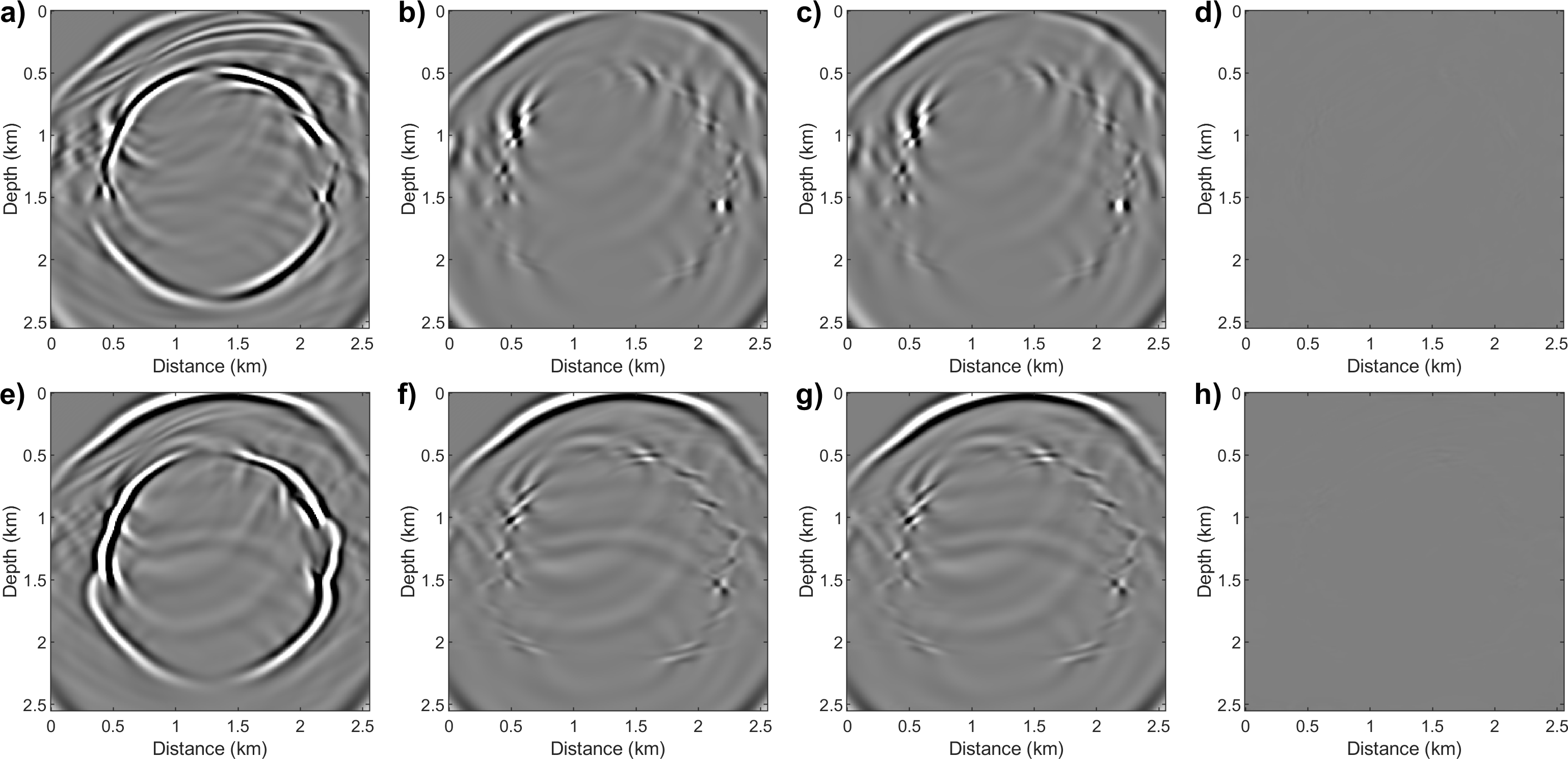}
\caption{Except for velocity models, the arrangement of other panels is the same as in Figure~\ref{fig7}, where we place the source in the model's interior center and record elastic wavefield at $t=0.6\,\mathrm{s}$. }
\label{fig8}
\end{figure*}

Again, we relocate the source to the model's interior center and analyze the recorded wavefields at 0.6 s. Figure~\ref{fig8} shows the corresponding separation results, maintaining the same panel structure as in Figure~\ref{fig7}, except for velocity models. Despite the altered source position and more challenging wave propagation caused by faults, our PICDM approach robustly and accurately extracts the P-wave modes. The residual plots show no significant energy leakage or shape discrepancy. This further demonstrates the effectiveness of our method in handling complex velocity structures and diverse source scenarios, confirming its strong generalization potential within the distribution. 

\subsection{Out-of-distribution models}
After evaluating our method on several in-distribution velocity models, we now shift our focus to assessing the generalization capability of our PICDM approach when applied to velocity models outside the training distribution. 

\begin{figure*}[htbp]
\centering
\includegraphics[width=1\textwidth]{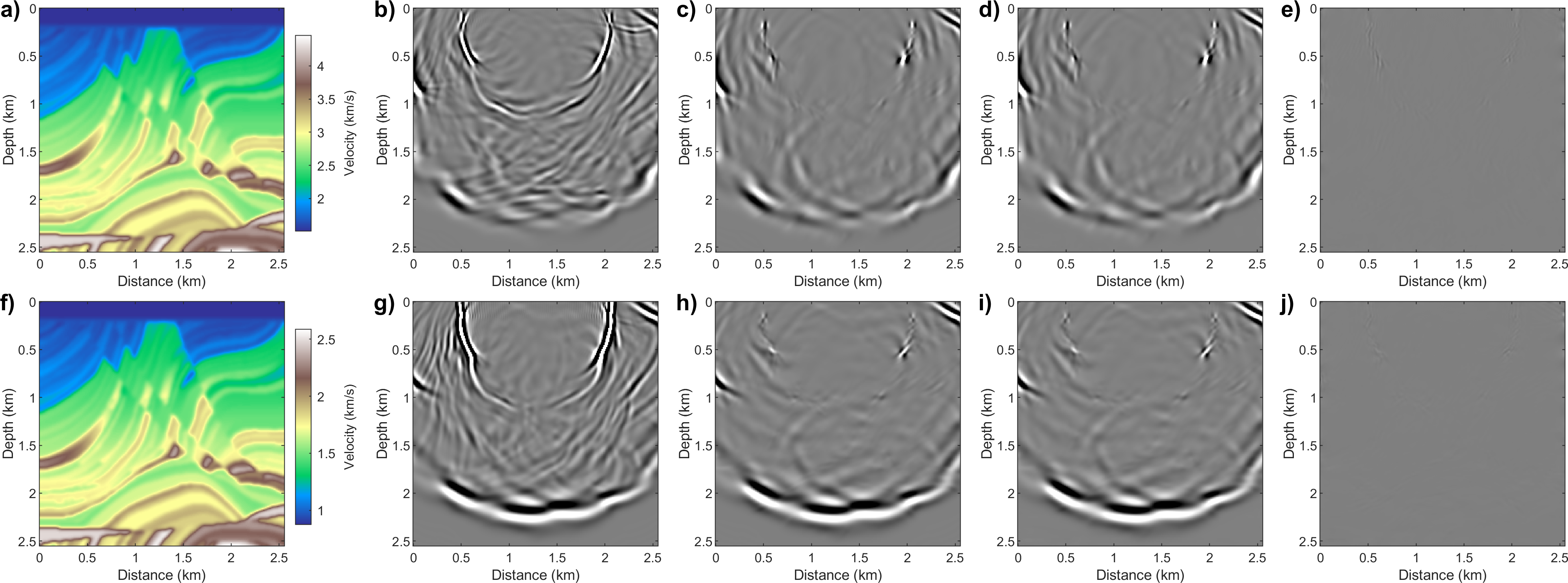}
\caption{Similar with Figure~\ref{fig5}, but for Marmousi 2 model. }
\label{fig9}
\end{figure*}

Initially, we select an area from the Marmousi 2 velocity model \citep{martin2006marmousi2}, which features complex fault structures. We construct the corresponding S-wave velocity model by multiplying the P-wave velocities by a factor of 0.58. The extracted P- and S-wave velocity profiles are presented in Figure~\ref{fig9}a and \ref{fig9}f, respectively. A source is placed at the top-center surface, and the elastic wavefield snapshot at 1.0 s is used for testing. 

Figure~\ref{fig9} (columns 2 to 5) displays the wavefield separation results on the Marmousi 2 model, showing the original elastic wavefields, the numerically separated P-wave modes, our model's predicted P-wave modes, and residuals between numerical and PICDM-predicted results. The top and bottom rows correspond to the horizontal ($x$) and vertical ($z$) velocity components, respectively. We can see that that our method consistently and accurately separates the P-wave mode, closely matching the numerical reference solution, demonstrating strong performance on out-of-distribution models.

\begin{figure*}[htbp]
\centering
\includegraphics[width=1\textwidth]{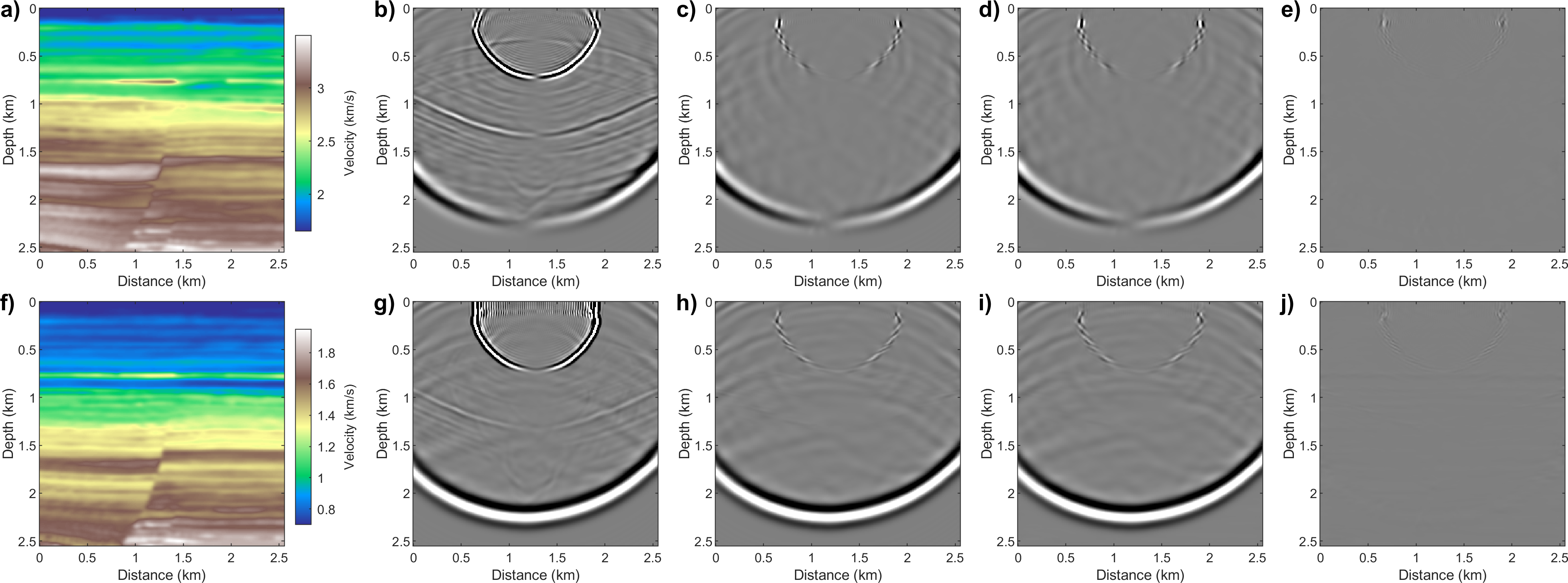}
\caption{Similar with Figure~\ref{fig5}, but for Otway model. }
\label{fig10}
\end{figure*}

Recognizing that real-world subsurface structures may exhibit complex, non-proportional relationships between P- and S-wave velocities, we further test our method using the realistic Otway velocity model, which contains distinct, independently measured P- and S-wave velocities. The extracted segments of these velocities are shown in Figures~\ref{fig10}a and \ref{fig10}f. Similar to the previous test, we position the seismic source at the center of the top surface and record the elastic wavefield at 1.0 s. Figure~\ref{fig10} presents our wavefield separation results for the Otway model, maintaining the same layout as in Figure~\ref{fig9}. Again, our PICDM demonstrates robust and precise separation of the P-wave modes, highlighting its capability to handle realistic and highly heterogeneous subsurface velocity distributions effectively.

\subsection{Evaluation on larger and realistic model sizes}
\begin{figure*}[htbp]
\centering
\includegraphics[width=1\textwidth]{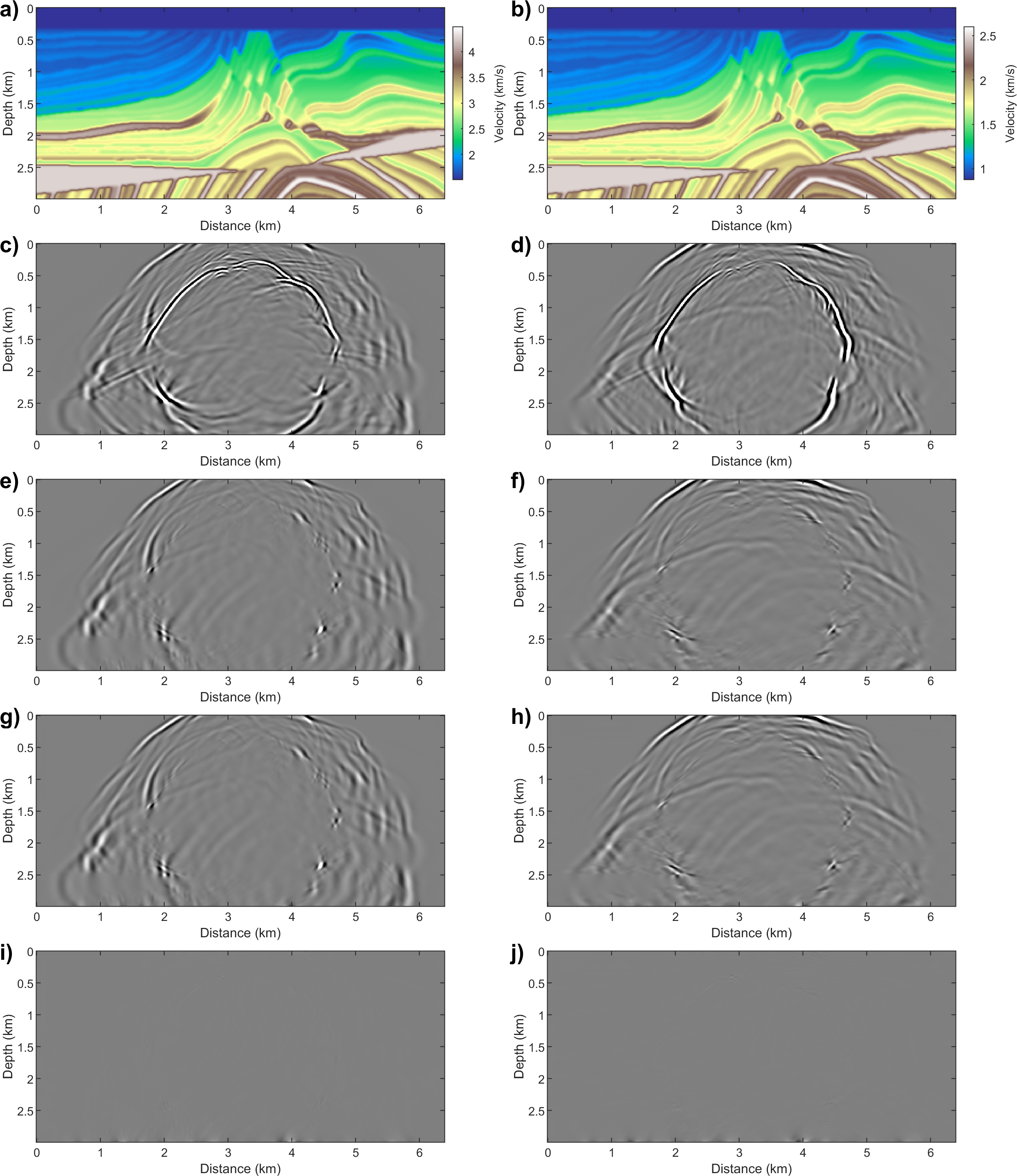}
\caption{Separation performance for a larger and more realistic Marmousi model (size \(300 \times 640\)) with the source placed at the center of the model at $t=1\,\mathrm{s}$. (a) P- and (b) S-wave velocity model. Rows 2 to 5 illustrate the wavefield separation outcomes, arranged from top to bottom as the original elastic wavefields, numerically separated P-wave modes (reference), our model's predicted P-wave modes, and residuals between numerical and our predictions. The left and right columns correspond to horizontal ($x$) and vertical ($z$) velocity components, respectively. All wavefields and residuals are presented on the same scale.}
\label{fig11}
\end{figure*}

In real-world seismic applications, velocity models typically are of larger dimensions compared to our training data. Thus, we now evaluate whether our PICDM, trained on models with dimensions \(256 \times 256\), can effectively generalize and upscale to larger and more realistic models. To this end, we modify the original Marmousi 2 model and transform it to a size of \(300 \times 640\) grid points. The P- and S-wave velocity profiles for the modified Marmousi 2 model are displayed in Figures~\ref{fig11}a and \ref{fig11}b. We place the source at the center of the model and record the elastic wavefield at 1.0 s for evaluation. 

Figure~\ref{fig11} (rows 2 to 5) illustrates our wavefield separation results for this larger Marmousi model. Rows from top to bottom show the original elastic wavefields, numerically separated P-wave modes, our model's predicted P-wave modes, and residuals between numerical and our predictions, respectively. The left and right columns correspond to horizontal ($x$) and vertical ($z$) components. The results clearly demonstrate that our PICDM method maintains excellent separation accuracy even on a significantly larger, out-of-distribution velocity model. The predicted P-wave modes closely match the numerical results, with only minimal energy leakage noticeable near the model's bottom boundary. These outcomes further confirm the robustness and scalability of our approach to realistic model sizes.

\subsection{Evaluation of temporal extrapolation}
To this point, all experiments have been conducted within the same 1.5 s time window used for training. We now investigate whether our PICDM framework can successfully extrapolate beyond this temporal range. Here, we revisit the modified Marmousi 2 model (see Figures \ref{fig11}a and \ref{fig11}b) used in previous subsection, place the source at the center of the top surface, and apply the approach on the elastic wavefield at 2.0~s for testing. 

\begin{figure*}[htbp]
\centering
\includegraphics[width=0.9\textwidth]{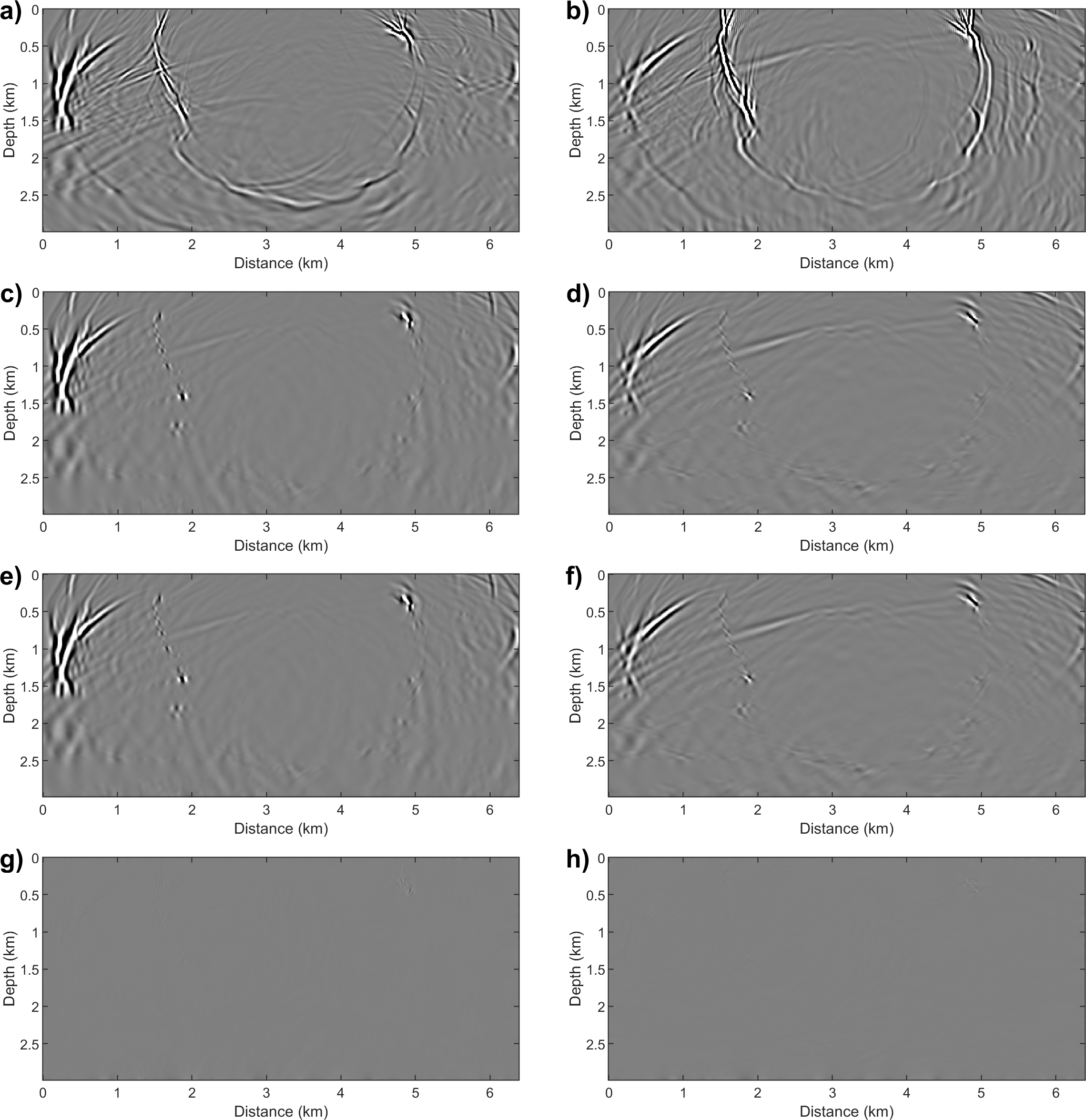}
\caption{Separation performance for the modified Marmousi 2 model (see Figures \ref{fig11}a and \ref{fig11}b) with the source placed at the middle of the top surface at $t=2.0\,\mathrm{s}$. Except for velocity models, the arrangement of other panels is the same as in Figure~\ref{fig11}. }
\label{fig12}
\end{figure*}

Figure~\ref{fig12} presents the P-wave mode separation results for this 2.0 s snapshot, using the same panel arrangement as in Figure \ref{fig11}, except that velocity model panels are omitted to avoid redundant display. We can see that, despite extrapolating in time beyond the training window, our method continues to accurately recover the P-wave mode, with residuals remaining negligible. This surprising result demonstrates the strong temporal extrapolation ability of the PICDM framework. This capability is critical for practical cases, as it allows our model to reliably handle time slices beyond the trained range, ensuring robust performance across extended temporal windows. 

\begin{figure*}[htbp]
\centering
\includegraphics[width=0.9\textwidth]{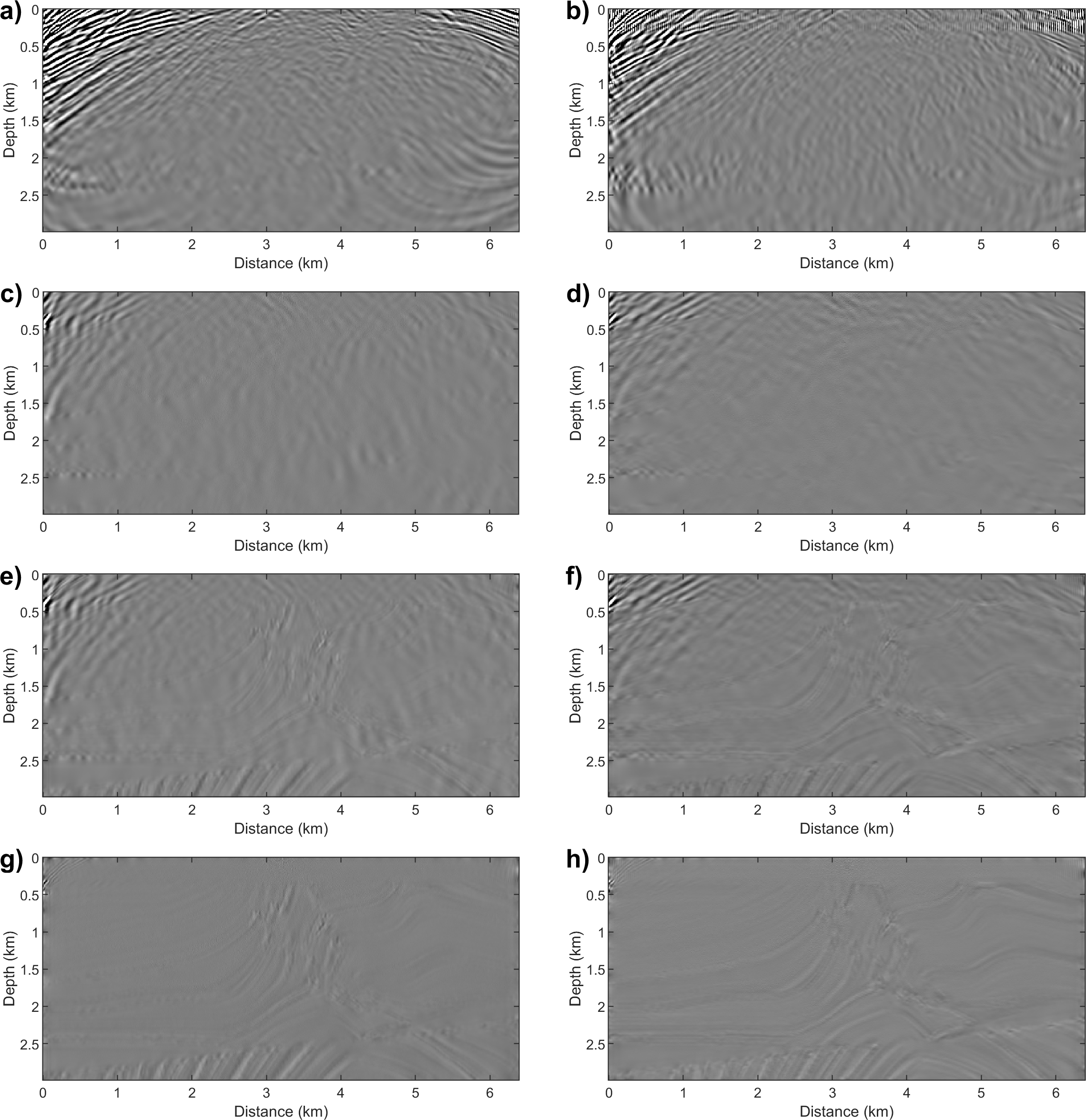}
\caption{Similar with Figure \ref{fig12}, but for snapshot at $t=4.0\,\mathrm{s}$. }
\label{fig13}
\end{figure*}

However, the temporal extrapolation capability of our trained diffusion model is not unlimited. For example, we present a snapshot of the elastic wavefield at 4.0 s for separation testing in Figure \ref{fig13}. The panel layout of Figure \ref{fig13} is identical to that of Figure \ref{fig12}. We can see that when the time extends significantly beyond the training window (1.5 s), the separated wavefields begin to exhibit artifacts tied to the velocity model. This result suggests that, while our PICDM can reliably handle moderate time beyond training scenarios, its performance is not unbounded in time. Training on a broader range of wavefield time slices would likely extend its effective extrapolation window and further strengthen its temporal generalization capability.

\section{\textbf{Discussion}}
The extensive numerical experiments demonstrated that our physics-informed conditional diffusion model (PICDM) generalizes remarkably well across a wide range of scenarios: from simple homogeneous media to complex in-distribution models (SEAM Arid and Overthrust), and from out-of-distribution cases (Marmousi 2 and Otway) to substantially larger model sizes (larger Marmousi). Remarkably, the method also maintains high accuracy when extrapolating temporally beyond the training window, highlighting its robustness in both spatial and temporal dimensions. Crucially, all of these results stem from a single training run. Our diffusion model is trained once and then applied across every test case with no additional retraining or fine-tuning. This learn-once, apply-everywhere paradigm underscores the PICDM’s powerful generalization capability and confirms that it delivers both high-fidelity P-wave separation and strong out-of-sample performance without any further training. 

This powerful generalization behavior suggests that the model has effectively learned the underlying separation operator described in Equation~\ref{eq2}. In essence, the diffusion network approximates the mapping from elastic wavefield to its P‑wave mode, replicating the action of the analytical decoupling operator. Similar to prior demonstrations of generative diffusion models serving as efficient neural operators in the frequency domain wavefield solutions \citep{cheng2025seismic}, PICDM internalizes the physics of wave propagation and separation, enabling rapid inference on previously unseen velocity configurations. 

\begin{figure*}[htbp]
\centering
\includegraphics[width=1\textwidth]{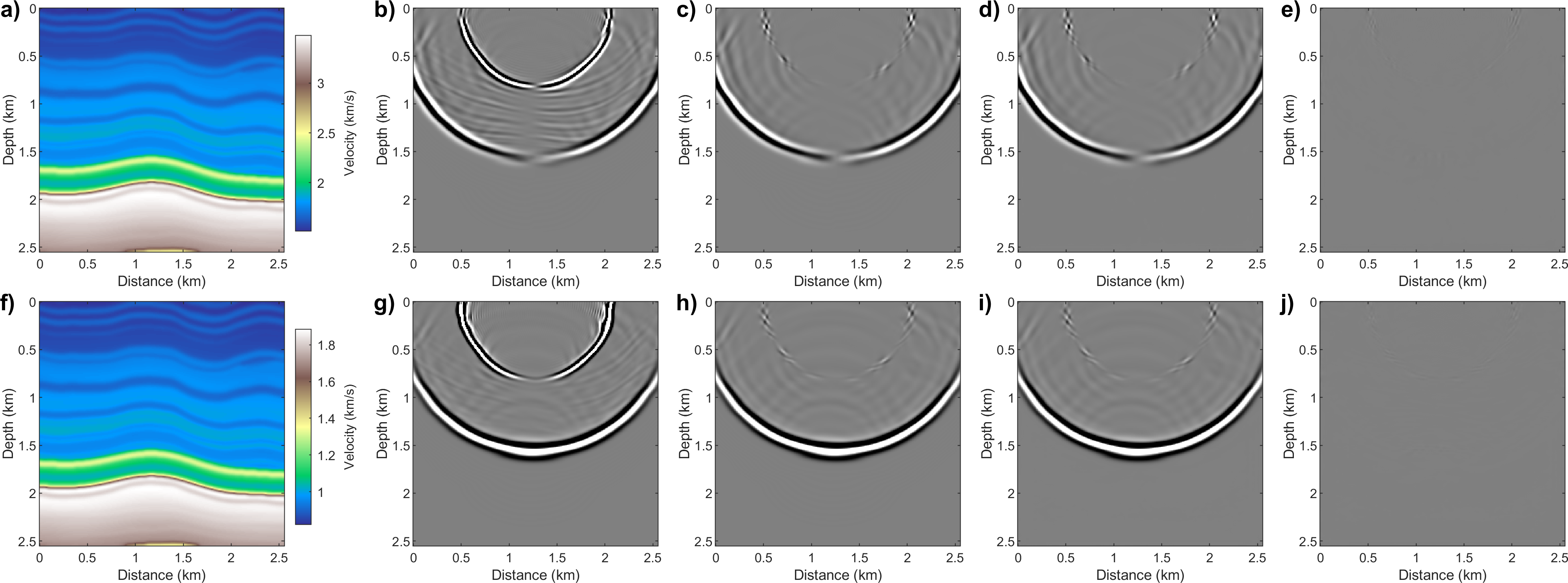}
\caption{Similar with Figure \ref{fig5}, but employing the velocity-free PICDM model.}
\label{fig14}
\end{figure*}

\begin{figure*}[htbp]
\centering
\includegraphics[width=1\textwidth]{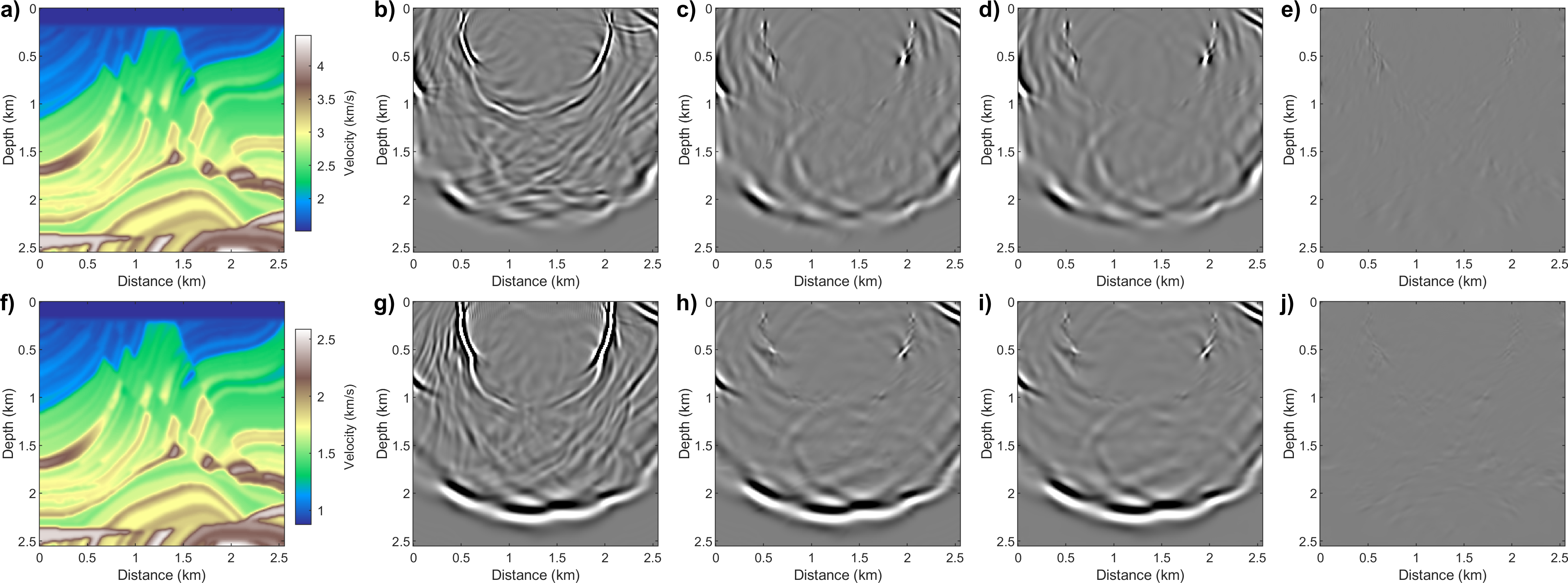}
\caption{Similar with Figure \ref{fig9}, but employing the velocity-free PICDM model.}
\label{fig15}
\end{figure*}

In fact, Equation~\ref{eq2} contains no explicit velocity-dependent terms. We therefore ask: can we remove the velocity model from the condition inputs $\mathbf{c}$ and still achieve reliable separation? To investigate, we retrain PICDM under identical hyperparameters and numerical setups, omitting the velocity input altogether. First, we evaluate this velocity-free PICDM on the in-distribution SEAM Arid model using the same elastic wavefields as in Figure \ref{fig5}. The results, shown in Figure \ref{fig14}, reveal that even without velocity conditioning, our PICDM produces separation accuracy nearly identical to the original velocity-conditioned version for in-distribution case. Next, we test the velocity-free PICDM on an out-of-distribution scenario by applying it to the extracted Marmousi velocity model and corresponding wavefield (see Figure \ref{fig9}). As displayed in Figure \ref{fig15}, the separation residuals are noticeably larger compared to the velocity-conditioned PICDM. This outcome implies that while velocity conditioning is not strictly necessary for in-distribution cases, it significantly enhances robust generalization when encountering unseen velocity structures. 

\begin{figure*}[htbp]
\centering
\includegraphics[width=1\textwidth]{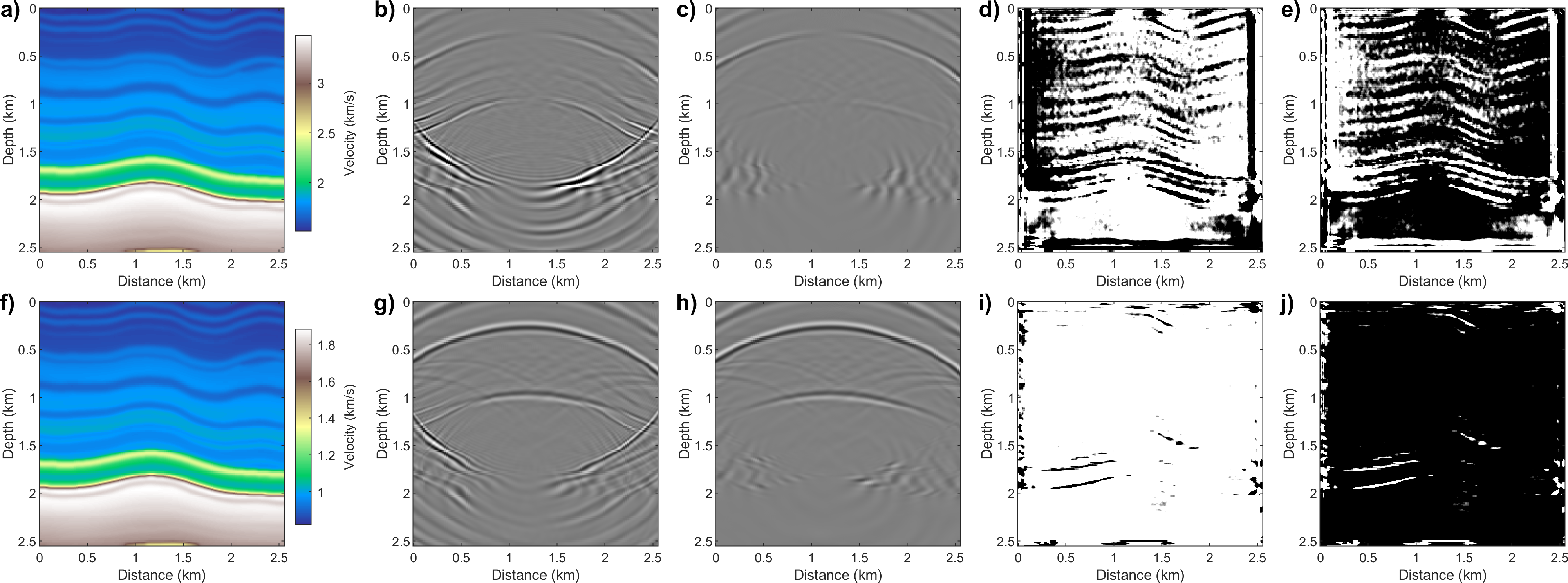}
\caption{Similar with Figure \ref{fig5}, but the peak frequency of the elastic wavefield is 18 Hz.}
\label{fig16}
\end{figure*}

However, our current study focused on a fixed source wavelet, in particular with a 12 Hz peak frequency. As shown in Figure~\ref{fig16}, when the PICDM model trained on 12 Hz data is applied to an elastic wavefield with an 18 Hz peak frequency, the separation fails, where the predictions are dominated by spurious artifacts. This outcome highlights a limitation of single-frequency training: the learned operator does not generalize across widely different frequency contents. Real seismic data encompass broad, often multimodal frequency spectra. Incorporating multi-peak-frequency training examples would allow the model to learn frequency-dependent aspects of the separation operator, potentially further enhancing generalization and accuracy across realistic seismic conditions. 

On the other hand, although we have addressed isotropic elastic wavefields, real subsurface media often exhibit anisotropy, where separation operators become considerably more complex and computationally expensive. Extending PICDM to anisotropic elastic wavefield separation is a promising direction, as the trained model could replace costly numerical implementations and accelerate high-resolution seismic imaging in anisotropic environments.

\section{\textbf{Conclusions}}
We presented a novel physics-informed conditional diffusion model for separating the P-wave mode from elastic wavefields. Unlike standard data-driven approaches, our method directly predicts the P-wave mode while incorporating physics-based constraints into both the training loss and the inference (reverse diffusion) process. This design allows the model to leverage not only the elastic wavefields and subsurface velocity models, but also the fundamental relationships governing elastic wave propagation. Comprehensive evaluations demonstrated that the proposed method achieves near-perfect agreement with traditional numerical methods for P-wave mode separation on diverse cases, spanning simple homogeneous media, complex in-distribution models (SEAM Arid, Overthrust), out-of-distribution cases (Marmousi 2, Otway), larger-scale models, and temporal extrapolation. Moreover, the physics-guided sampling strategy accelerates convergence and reduces residual errors, enabling efficient inference with fewer diffusion steps. These results suggest that our PICDM framework generalizes well across spatial scales, velocity structures, and time windows. As a result, our method offers an effective and physics-grounded solution for elastic wavefield separation.
\section{\textbf{Acknowledgment}}
We thank Mohammad H. Taufik for providing the velocity model dataset used in our experiments. We also gratefully acknowledge the authors of the improved denoising diffusion probabilistic models open-source implementation \citep{nichol2021improved}, upon which our diffusion model architecture and code development were built. This publication is based on work supported by the King Abdullah University of Science and Technology (KAUST). The authors thank the DeepWave sponsors for their support. This work utilized the resources of the Supercomputing Laboratory at King Abdullah University of Science and Technology (KAUST) in Thuwal, Saudi Arabia. 
\section{\textbf{Code Availability}}
The data and accompanying codes that support the findings of this study are available at: 
\url{https://github.com/DeepWave-KAUST/DiffSeparation}. (During the review process, the repository is private. Once the manuscript is accepted, we will make it public.)

\bibliographystyle{unsrtnat}
\bibliography{references}

\end{document}